\documentclass[aps, twocolumn, showpacs, preprintnumbers, superscriptaddress, amsmath, amssymb, prb]{revtex4}
\usepackage{graphicx}
\usepackage{dcolumn}
\usepackage{bm}
\usepackage{color}
\usepackage{amsfonts}

\begin{document}

\title{
Dynamical Spectral Function From Numerical Renormalization Group: A Full Excitation Approach }

\author{Ke Yang}
\affiliation{Department of Physics, Renmin University of China, 100872 Beijing, China}
\author{Ning-Hua Tong}
\email{nhtong@ruc.edu.cn}
\affiliation{Department of Physics, Renmin University of China, 100872 Beijing, China}
\date{\today}

\begin{abstract}
For a given quantum impurity model, Wilson's numerical renormalization group (NRG) naturally defines a NRG Hamiltonian whose exact eigenstates and eigenenergies are obtainable. We give exact expressions for the free energy, static, as well as dynamical quantities of the NRG Hamiltonian. The dynamical spectral function from this approach contains full excitations including intra- and inter- shell excitations. For the spin-boson model, we compare the spectral function obtained from the present method and the full density matrix (FDM) method, showing that while both guarantee rigorous sum rule, the full excitation approach avoids the causality problem of FDM method. 
\end{abstract}



\maketitle
\section{Introduction}

Wilson's numerical renormalization group (NRG) method\cite{Wilson1,Bulla3} is powerful for studying quantum impurity models. Since its invention, NRG has witnessed a series of development, including z-averaging to mitigate the discretization error,\cite{Yoshida1} improvements in the logarithmic discretization,\cite{Campo1,Zitko1} extension to bosonic systems,\cite{Bulla1} full density matrix algorithm,\cite{Weichselbaum1} extension to time dependence,\cite{Anders1} and merging with matrix product state\cite{Saberi1,Weichselbaum3,Pizorn1} and tensor network,\cite{Weichselbaum2} etc. Today, both the sophistication and applicability of NRG have been advanced significantly compared to Wilson's original work. 

The calculation of spectral function of an quantum impurity model from the NRG-produced eigenstates and eigenenergies is an important problem. The patching method\cite{Bulla2} combines the spectral functions from successively lower energy shells to produce a full spectral function which does not guarantee the exact sum rule. Using the reduced density matrix of the full system to combine the spectral functions of different energy shells, Hofstetter\cite{Hofstetter1} developed the density-matrix NRG that can take int account the influence of the low energy states to the high frequency spectral function. In the full density matrix (FDM) NRG method,\cite{Weichselbaum1} the Lehmannn representation of the spectral function is treated with a complete set of eigenstates and simplified by the NRG approximation. FDM NRG fulfils the sum rule rigorously and accurately describes the spectral features at energies below the temperature. Now, FDM method is the most widely used method for producing spectral functions of quantum impurity models within NRG.

Although the FDM method is highly accurate and efficient in general, in this paper, we illustrate that FDM has a problem of causality which, in certain situations, leads to negative spectral functions. This problem arises from the approximate treatment of the unitary time evolution of operators in the Green's function by the NRG approximation used in FDM. With this approximation, the excitations between different NRG shells  (inter-shell excitations) are approximated by the kept-discarded excitations within each NRG shells (intra-shell excitations). We demonstrate this problem using the spin-boson model (SBM) in the parameter regime of strong coupling and finite bias. To circumvent this problem of FDM NRG, first, we point out that the algorithm of NRG naturally defines an effective projective Hamiltonian $\tilde{H}_{N}$, dubbed NRG Hamiltonian. The complete basis proposed by Anders et al.\cite{Anders1} is the set of exact eigenstates of $\tilde{H}_{N}$, with their eigenenergies being generated by NRG calculation. Then, we propose an algorithm to calculate the exact free energy, static, as well as dynamical quantities of $\tilde{H}_{N}$, which constitute well-controlled approximations to those of the original impurity model. The obtained spectral function contains both the intra- and the inter- shell excitations. It satisfies the rigorous sum rule and positiveness. Hereafter this new algorithm is called full excitation (FE) NRG method.

\section{FE formalism}
In this section, we derive the formalism of FE method for general quantum impurity models.
The Hamiltonian of a generic quantum impurity model reads $H = H_{imp} + H_{bath} + H_{c}$.
A small quantum system described by $H_{imp}$ is coupled through $H_c$ to a continuous non-interacting reservoir described by $H_{bath} = \sum_{i} \epsilon_i c_{i}^{\dagger}c_{i}$. Here, $c_{i}^{\dagger}$ creates a particle (fermion or boson) with energy $\epsilon_i$. The indices such as spin and orbital are included in $i$. The impurity is coupled directly to the local bath degrees of freedom $f_{0} = 1/L \sum_i V_i c_i$, with $L$ the normalization constant.

The NRG algorithm consists of three steps.\cite{Wilson1,Bulla3} (i) The continuous bath degrees of freedom are discretized into bath sites with exponentially descending energies $\omega_n \sim \Lambda^{-n}$. $\Lambda \geqslant 1.0$ is the logarithmic discretization parameter. This step introduces the logarithmic discretization error which diminishes as $\Lambda$ decreases to unity. (ii) The discretized Hamiltonian is canonically transformed into a semi-infinite chain of the form (truncated to length $N$ and neglecting possible indices of spin, orbital, etc.)
\begin{eqnarray}
  H_{N} &=& H_{imp} + c_0 \left(f_0^{\dagger} A + A^{\dagger} f_0 \right)
  + \sum _{n=0}^{N} \epsilon_n f_n^{\dagger} f_n   \nonumber \\
  &&  + \sum_{n=0}^{N-1} t_n \left( f_{n}^{\dagger} f_{n+1} + f_{n+1}^{\dagger} f_{n} \right) .
\end{eqnarray}    \label{Eq.1}
Here, $A$ is an impurity operator. $H_{N}(\Lambda)$ is a function of $\Lambda$. Both $\epsilon_n$ and $t_n$ decay as $\Lambda^{-n/2}$ for fermionic bath ($\Lambda^{-n}$ for bosonic bath). 
(iii) The chain Hamiltonian is diagonalized iteratively. Starting from the longest chain $H_{n_0}$ whose all eigenstates can be kept, we add one bath site and diagonalize the enlarged system. This is done iteratively until all the chain sites are added and diagonalized. To handle the divergence of the Hilbert space in this process, after diagonalizing $H_{n}$, only the $M$ eigenstates with lowest eigenenergies are kept. The matrix of $H_{n+1}$ is built in the product space of these kept states and the bare states of the newly added site. The truncation error introduced in this step diminishes in the limit $M=\infty$.

\begin{figure}[t!]
\vspace{-5.0cm}
\begin{center}
\includegraphics[width=5.8in, height=3.6in, angle=0]{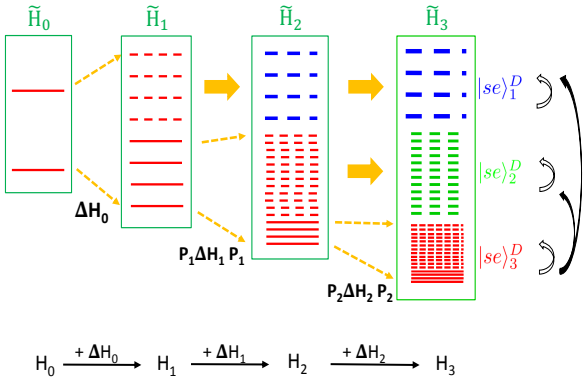}
\vspace*{-1.0cm}
\end{center}
\caption{Schematic picture for the structure of the spectrum of $\tilde{H}_n$. We suppose $H_0$ contains only two levels, keep $M=4$ lowest states, and take the Hilbert space dimension of each bath site $d=4$ and chain length $N=3$. The upper part shows what is practically done in NRG calculation. Solid horizontal levels denote the kept states. Dashed horizontal levels denote the discarded states with degeneracies proportional to the thickness of line. The dashed thin arrows show the splitting of the kept energy levels by adding $P_i \Delta H_i P_i$ to $H_i$ ($i=0,1,2$). Here $P_i$ is the projecting operator of the kept subspace of $\tilde{H}_i$. The wide arrows show the $4$-fold expanding of degeneracy on adding a bath site. The solid rectangulars mark out the complete eigen spectrum of $\tilde{H_i}$ ($i=0 \sim 3$). $\tilde{H}_2 = \tilde{H}_{1} + P_1 \Delta H_1 P_1$ and $\tilde{H}_3 = \tilde{H}_{2} + P_2 \Delta H_2 P_2$. Note for the beginning short chain $H_{n_0=0}$ whose all eigenstates are kept, $\tilde{H}_{0} = H_0$. $\tilde{H}_1 = \tilde{H}_{0} + P_0 \Delta H_0 P_0 = H_1$ because $P_0= 1$. The intra- and inter-shell excitations are shown by empty and solid arrows on the right side, respectively. In the bottom are the  theoretical equations where the truncation is not taken into account explicitly.
} \label{Fig1}
\end{figure}

NRG calculation generates many eigenstates $|s \rangle_n$ and eigenenergies $E_{ns}$ \cite{Note1} ($n \in [n_0, N]$, $s \in [1, D_n]$). Here $D_n$ is the number of produced eigenstates by diagonalizing $H_n$. For each $n$, the lowest $M$ states are kept and the higher $D_n - M$ ones are discarded. They are denoted as $|s \rangle_n^{K}$ and $|s \rangle_n^{D}$, respectively. For the last shell $n=N$, all the states are regarded as discarded.

Let us analyse the structure of the eigen spectrum generated by NRG. Suppose $H_{n+1} = H_n + \Delta H_n$ ($n=0,1,...,N-1$). $\Delta H_n$ contains the on-site energy of the newly added bath site $n+1$ and the hopping between sites $n+1$ and $n$. If $\Delta H_n$ is not considered, adding bath site $n+1$ will increase the degeneracy of each eigenstate of $H_{n}$ by a factor of $d$. If $\Delta H_n$ is fully added, all these degeneracies will be lifted. In NRG calculation, $\Delta H_n$ is added only partly. That is, the matrix of $\Delta H_n$ is constructed in the space of kept states of $H_{n}$ multiplying the bare states of bath site $n+1$. Therefore, the degeneracies in the extended spectrum of $\tilde{H}_n$ are partly lifted. The resulting Hamitonian matrix corresponds to the Hamiltonian 
\begin{equation}
   \tilde{H}_{n+1} = \tilde{H}_{n} + P_n \Delta H_{n} P_n
\end{equation}       \label{Eq.2}
instead of to the theoretical $H_{n+1} = H_n + \Delta H_{n}$. Here $P_n$ is the projecting operator of the kept space of $\tilde{H}_n$.  The full spectrum of $\tilde{H}_{n+1}$ (rectangular boxes in Fig.1 for $n=1$ and $2$) is composed of those low energy eigenstates obtained from lifting the degeneracies of $\tilde{H}_{n}$ by $P_n \Delta H_n P_n$ (red horizontal levels in Fig.1), and those high energy eigenstates generated by multiplying new bath states to previous eigenstates while maintaining degeneracies (green and blue horizontal levels in Fig.1). 
A schematic picture is shown in Fig.1 for illustration of the above process, using the number of kept states $M=4$, Hilbert space dimension of bath site $d=4$, and chain length $N=3$. Detailed explanation is in figure caption.

Grouping all the discarded states (extended to include the degeneracies) generated in the calculation for $H_N$ (e.g., the rectangular box of $H_3$ in Fig.1), we obtain not only a complete basis set for $H_N$,\cite{Anders1} but also the exact eigenstates of the following {\it NRG Hamiltonian} $\tilde{H}_{N}$ 
\begin{eqnarray}
  \tilde{H}_{N} &=& H_{n_{0}} + \sum_{n=n_{0}+1}^{N} \epsilon_{n} f_{n}^{\dagger} f_{n}  \\
    && + \sum_{n=n_{0}}^{N-1}  t_n  \left[ \left(P_n f_{n}^{\dagger}P_n \right) f_{n+1} + f_{n+1}^{\dagger} \left(P_n f_{n} P_n \right) \right], \nonumber 
\end{eqnarray}    \label{Eq.3}   
which is obtained by iterating Eq.(2) and setting $\tilde{H}_{n_0} = H_{n_0}$.  $\tilde{H}_{N} =\tilde{H}_{N}(\Lambda, M) $ depends on the NRG parameter $\Lambda$ and $M$. It is a many-body Hamiltonian defined in the original Hilbert space of $H_N$, with eigenstates $| se \rangle^{D}_{n}$ and eigenenergies $E^{D}_{ns}$ ($n \in [n_0+1, N]$),
\begin{equation}
   \tilde{H}_{N} |s e \rangle_{n}^{D} = E_{ns}^{D}|s e \rangle_{n}^{D}.
\end{equation}    \label{Eq.4}
Here, $|s e \rangle_{n}^{D} = | e_n \rangle \otimes | s \rangle_{n}^{D} = | \sigma_N \sigma_{N-1} ... \sigma_{n+1} \rangle \otimes | s \rangle_{n}^{D}$ are matrix product states with degeneracy $d^{N-n}$. $|\sigma_{n}\rangle$'s ($\sigma_n \in [1, d]$) are the bare basis states of site $n$. The eigenstates fulfil the standard orthonormal and complete relations.\cite{Anders1,Weichselbaum1} We have $H = H_{N=\infty}(\Lambda=1.0)$ and $H_{N}(\Lambda) = \tilde{H}_{N}(\Lambda, M=\infty)$. Therefore, $\tilde{H}_{N}$ approximates $H_N$ of Eq.(1) with the control parameter $M$ and $H_N$ approximates $H$ with the control parameter $\Lambda$. Thanks to the exponential separation of energy scales due to the logarithmic discretization and the truncation scheme of NRG, $\tilde{H}_{N}$ has very accurate low energy states. 
Note that the extended kept states $|se \rangle^{K}_{n}$ ($n=0,1,...,N-1$) are not exact eigenstates of $\tilde{H}_{N}$. 

We now consider to produce the exact physical quantities of $\tilde{H}_{N}$ from the obtained $\{ |s \rangle_n \}$ and eigenenergies $\{ E_{ns} \}$.
The partition function $Z$ at temperature $T$ reads
\begin{equation}
 Z =  \sum_{n=n_0+1}^{N} \sum_{s} d^{N-n} e^{-\beta E_{ns}^{D} } .
\end{equation}    \label{Eq.5}
The exact free energy of $\tilde{H}_{N}$ is $F=-(1/\beta)\ln{Z}$.
The statistical average of an impurity operator $\hat{O}$ reads
\begin{equation}
   \langle \hat{O} \rangle = \frac{1}{Z} \sum_{n=n_0+1}^{N} \sum_{s} d^{N-n} e^{-\beta E_{ns}^{D} } \, {^D_n}\langle s | \hat{O} |s \rangle_n^{D}.
\end{equation}    \label{Eq.6}

The above expressions were already employed in the FDM method which treats the density matrix exactly.\cite{Weichselbaum1,Weichselbaum2}
FE and FDM differ in their formalisms for dynamical quantities. Consider, for example, the time correlation function $\langle A(t) B \rangle = Tr \left[\rho A(t)B \right]$ of two impurity operators $A$ and $B$. The density operator reads $\rho = e^{-\beta H}/Z$. Inserting $1 = \sum_{nse} |se\rangle_{n}^{D} {^D_n}\langle se|$ twice, we obtain 
\begin{eqnarray}
&& \langle A(t) B \rangle = \nonumber \\
&& \sum_{nse} \sum_{n^{\prime}s^{\prime}e^{\prime} } Tr \left[ \rho |se\rangle^{D}_n {^D_n}\langle se| e^{iHt}Ae^{-iHt}  |s^{\prime}e^{\prime} \rangle^{D}_{n^{\prime}} {^D_{n^{\prime}}} \langle s^{\prime}e^{\prime} | B \right].  \nonumber \\
&&
\end{eqnarray}      \label{Eq.7}
Once the exact expression Eq.(4) is used to evaluate the matrix elements of $A(t)$, excitations of the form $E^{D}_{ns} - E^{D}_{n^{\prime} s^{\prime}}$ will be generated, which include both inter- ($n \neq n^{\prime}$) and intra-shell ($n = n^{\prime}$) excitations. Before we present the FE formalism, we first make a briefly analysis for the FDM method.\cite{Weichselbaum1} To obtain the formalism of FDM, we first reduce Eq.(7) into the single-shell form with the help of the exact relation 
$\mathbf{1}^{d_0 d^{N-n_0}} = \sum_{se} |se\rangle^{K}_m {^K_m}\langle se| + \sum_{n=n_0+1}^{m} \sum_{se} |se\rangle^{D}_n {^D_n}\langle se|$ \,\, ($m \in [n_0, N]$). We obtain
\begin{eqnarray}
&& \langle A(t) B \rangle = \nonumber \\
&& \sum_{n=n_0+1}^{N} \sum_{se s^{\prime}e^{\prime} } Tr \left[ \rho |se\rangle^{D}_n {^D_n}\langle se| e^{iHt}Ae^{-iHt}  |s^{\prime}e^{\prime} \rangle^{D}_{n} {^D_{n}} \langle s^{\prime}e^{\prime} | B \right]  \nonumber \\
&& + \sum_{n} \sum_{se s^{\prime}e^{\prime} } Tr \left[ \rho |se\rangle^{D}_n {^D_n}\langle se| e^{iHt}Ae^{-iHt}  |s^{\prime}e^{\prime} \rangle^{K}_{n} {^K_{n}} \langle s^{\prime}e^{\prime} | B \right]  \nonumber \\
&& + \sum_{n} \sum_{se s^{\prime}e^{\prime} } Tr \left[ \rho |se\rangle^{K}_n {^K_n}\langle se| e^{iHt}Ae^{-iHt}  |s^{\prime}e^{\prime} \rangle^{D}_{n} {^D_{n}} \langle s^{\prime}e^{\prime} | B \right]. \nonumber \\
&&
\end{eqnarray}      \label{Eq.8}
The exact relation $\rho |se\rangle_{n}^{D} = e^{-\beta E_{ns}^{D}}/Z$ is used for the density operator $\rho$. To calculate the matrix elements of $e^{iHt}Ae^{-iHt}$ in the second and third terms, the NRG approximation $H_{N} |se \rangle^{K}_n \approx E^{K}_{ns} |se \rangle^{K}_n$ is used on one side of $A$ and the exact Eq.(4) is used on the other side. The $e^{\pm iHt}$ factors on two sides of $A$ are hence not treated on equal footing. The obtained expression reads
\begin{eqnarray}
&& \langle A(t) B \rangle =  \nonumber \\
&& \sum_{n s s^{\prime} \tilde{s} }  \left[ B_{DD}^{(n)} \right]_{s^{\prime} \tilde{s}}  \left[\rho_{DD}^{(n)} \right]_{\tilde{s} s} \left[ A_{DD}^{(n)} \right]_{ss^{\prime}} e^{i \left( E^{D}_{ns} - E^{D}_{ns^{\prime}}\right)t} d^{N-n} \nonumber \\
&& + \sum_{n s s^{\prime} \tilde{s} } \left[ B_{KD}^{(n)} \right]_{s^{\prime} \tilde{s}}  \left[\rho_{DD}^{(n)} \right]_{\tilde{s} s} \left[ A_{DK}^{(n)} \right]_{ss^{\prime}} e^{i \left( E^{D}_{ns} - E^{K}_{ns^{\prime}}\right)t} d^{N-n}   \nonumber \\
&& + \sum_{n s s^{\prime} \tilde{s} } \left[ B_{DK}^{(n)} \right]_{s^{\prime} \tilde{s}}  \left( \sum_{e} {^{K}_n}\langle \tilde{s}e| \rho | se \rangle^{K}_{n} \right) \left[ A_{KD}^{(n)} \right]_{ss^{\prime}} e^{i \left( E^{K}_{ns} - E^{D}_{ns^{\prime}}\right)t}  \nonumber \\
&&
\end{eqnarray}      \label{Eq.9}
Here, the matrix elements are defined as $\left[ O_{XX^{\prime}}^{(n)} \right]_{ss^{\prime}} = {^{X}_{n}} \langle se|  O | s^{\prime} e \rangle^{X^{\prime}}_{n}$, for $O = A$, $B$, and $\rho$. Eq.(9) contains only intra-shell excitations among which the kept-discarded excitations are approximate for $\tilde{H}_{N}$.  For the case $B=A^{\dagger}$, the first two terms have positive weights since the matrix $\rho_{DD}^{(n)}$ is diagonal and positive. In the third term, the matrix $\rho_{KK}^{(n)}$ is not diagonal because the Hamiltonian of the full chain causes overlap between different kept states of the same iteration. For fixed $n$, $s$ and $s^{\prime}$, the prefactor of $\exp \left[i \left( E^{K}_{ns} - E^{D}_{ns^{\prime}}\right)t \right]$ is not guaranteed to be positive unless the same NRG approximation is used for $\rho_{KK}^{(n)}$. In summary, in the FDM formalism, the density operator $\rho$ is evaluated exactly but the matrix elements of $A(t)$ are treated with the NRG approximation. The inter-shell excitations are approximately replaced by the kept-discarded intra-shell excitations. As a result, albeit the spectral function fulfils the rigorous sum rule, the positiveness of the diagonal spectral function is lost.

In contrast, in deriving the FE formalism, we start from Eq.(7) and use Eq.(4) only. The obtained expression for the spectral function is exact for $\tilde{H}_{N}$. Naturally, it has no causality problem. 
Below, we focus on the retarded Green's function (GF) of the impurity operators $A$ and $B$, $G^{f/b}_{A,B}(\omega) \equiv -i \int_{0}^{+\infty} \langle \left[ A(t), B(t^{\prime}) \right]_{\pm}\rangle e^{i(\omega+ i\eta)(t-t^{\prime})} d(t-t^{\prime})$. $G^f$ ($G^b$) denotes Fermi- (Bose-) type GF which is defined with anti-commutator (commutator). Starting from the exact Lehmann representation of $G^{f/b}_{A,B}(\omega)$, inserting the complete relation of the complete basis, and using Eq.(4) to compute the matrix elements of both $\rho$ and $e^{\pm iHt}$, we obtain the FE formula for GF. Details of the derivation are summarized in Appendix. We obtain
\begin{equation}
 G^{f/b}_{A,B}(\omega) = \sum_{m=n_0+1}^{N} \sum_{s^{\prime}} w_{ms^{\prime}} G_{A,B}^{(ms^{\prime})}(\omega),
\end{equation}    \label{Eq.10}
with the weight $w_{ms^{\prime}} = (1/Z) d^{N-m} e^{-\beta E^{D}_{ms^{\prime}}}$ and 
\begin{eqnarray}
&& G_{A,B}^{(ms^{\prime})}(\omega) =  \nonumber \\
&& \sum_{s} \left[1 \pm e^{-\beta (E^{D}_{m s} - E^{D}_{ms^{\prime}}) }\right] \frac{\left[ A^{(m)}_{DD} \right]_{ss^{\prime}} \left[ B^{(m)}_{DD} \right]_{s^{\prime}s} }{\omega + i \eta + E^{D}_{ms} - E^{D}_{ms^{\prime}} }   \nonumber \\
&& + \sum_{n = n_0+1}^{m-1} \sum_{s}\left[1 \pm e^{-\beta (E^{D}_{n s} - E^{D}_{ms^{\prime}}) }\right] \frac{\left[ A B \right]^{(nm)}_{ss^{\prime}} }{\omega + i \eta + E^{D}_{ns} - E^{D}_{ms^{\prime}} } \nonumber \\
&& +  \sum_{n = n_0+1}^{m-1} \sum_{s}\left[1 \pm e^{-\beta (E^{D}_{n s} - E^{D}_{ms^{\prime}}) }\right] \frac{\left[ B A \right]^{(nm)}_{ss^{\prime}}  }{\omega + i \eta + E^{D}_{ms^{\prime}} - E^{D}_{ns} }. \nonumber \\
&&
\end{eqnarray}    \label{Eq.11}
In the equation, $\left[O^{(m)}_{DD} \right]_{ss^{\prime}} = {^D_m}\langle s| O |s^{\prime} \rangle_m^{D}$, ($O =A$,$B$). $\left[ XY \right]^{(nm)}_{ss^{\prime}}$ ($XY=AB$ or $BA$) is given by (for $n_0+1 \leq  n \leq m-1$)
\begin{equation}
 \left[ XY \right]^{(nm)}_{ss^{\prime}} = \sum_{s_1, s_2} \left[ X^{(n)}_{DK} \right]_{s s_1}  V^{(n,m)}_{s_1,s^{\prime}, s_2} \left[ Y^{(n)}_{KD} \right]_{s_2 s} .
\end{equation}    \label{Eq.12}

The transition matrix ${\bf V}$ is calculated recursively through
\begin{equation}
  V^{(n-1,m)}_{s_1, s^{\prime}, s_2} = \sum_{\sigma_n, s_3, s_4} \left[ U_{KK}^{(\sigma_n)}\right]_{s_1 s_3}   V^{(n,m)}_{s_3,s^{\prime}, s_4} \left[ U_{KK}^{(\sigma_n)}\right]^{\dagger}_{s_4 s_2} ,
\end{equation}    \label{Eq.13}
with the initial value 
\begin{eqnarray}
 V^{(m-1,m)}_{s_1,s^{\prime}, s_2} = \sum_{\sigma_m} \left[ U_{KD}^{(\sigma_m)}\right]_{s_1 s^{\prime}}   \left[ U_{KD}^{(\sigma_m)}\right]^{\dagger}_{s^{\prime} s_2}.
\end{eqnarray}    \label{Eq.14}
Here, the matrices $U_{KK}^{(\sigma_m)}$ and $U_{KD}^{(\sigma_m)}$ are respectively the kept-kept and kept-discarded blocks of the unitary transformation matrices produced by the diagonalization of $H_m$ in NRG.
Eqs.(10)-(14) are the main results of this paper. 
This FE formalism contains both intra- and inter-shell excitations. The unitarity of quantum evolution in $A(t)$ is maintained at the expense of introducing inter-shell excitations of $\tilde{H}_{N}$. For $B=A^{\dagger}$, Eq.(11) has a hermitian symmetry and naturally guarantees the positiveness of the spectral function $-(1/\pi) {\text Im}{G^{f/b}_{A,A^{\dagger}}(\omega)}$ in $\omega > 0$.

\section{Results and Comparison}

Below, we use the spin-boson model (SBM) \cite{Leggett1} to demonstrate FE algorithm and to make comparison with the patching method and FDM method. SBM describes a two-level quantum system coupled to a dissipative bosonic bath. It has been widely studied in many contexts ranging from superconducting qubit \cite{Makhlin1} to photosynthetic biosystems.\cite{Muehlbacher1} NRG has played an important role in the understanding of this model.\cite{Bulla1,Vojta1,Guo1,Tong1} The Hamiltonian reads 
\begin{equation}
   H_{SB} = -\frac{\Delta}{2} \sigma_x + \frac{\epsilon}{2}\sigma_z + \sum_i \omega_i a_i^{\dagger}a_i + \frac{\sigma_z}{2}\sum_{i}\lambda_i \left(a_i^{\dagger}+a_i \right). 
\end{equation}    \label{Eq.15}
The two-level system is described by Pauli matrices and the influence of bath is encoded into the spectral function $J(\omega) = \pi \sum_i \lambda_i^{2} \delta(\omega - \omega_i)$, for which we use $J(\omega) = 2 \pi \alpha \omega^{s} \omega_c^{1-s}$ ($0 < \omega < \omega_c$, $\omega_c=1.0$) with coupling strength $\alpha$ and exponent $s$. As usual, we truncate the Hilbert space of each boson site to $N_b$ states in the occupation basis.\cite{Bulla1} Now, the NRG Hamiltonian becomes $\tilde{H}_{N}(\Lambda, M, N_b)$ and FE method produces the exact quantities for it.
In this paper, we study the Fourier transform of the anti-symmetric dynamical correlation function,
\begin{eqnarray}
C(\omega) &=& \frac{1}{2\pi} \int_{-\infty}^{\infty} (1/2)\langle \left[ \sigma_{z}(t), \sigma_z(0) \right]_{+}\rangle e^{i\omega t} dt  \nonumber \\
 &=& -\frac{1}{2\pi} \text{Im} G^{f}_{\sigma_z \sigma_z}(\omega).
\end{eqnarray}    \label{Eq.16}

In Fig.2(a), we plot the regular part of $C(\omega)$ \cite{Note3} obtained from the patching method, FDM method, and FE for a sub-Ohmic bath $s=0.3$ at $\alpha > \alpha_c$, $\epsilon > 0$, and a low temperature. In this paper, we use the standard log-Gaussian broadening for the spectral function at all frequencies, being different from the fermion case where a Lorentzian broadening is used instead for $\omega < T$.\cite{Bulla2} The broadening is controlled by the width $B$ of the log-Gaussian function. The curve from FDM method agrees well with that of FE in the frequency regime $\omega/\Delta \gtrsim 10^{-3}$ but becomes negative in lower frequencies. Both curves fulfil the sum rule $\int_{-\infty}^{\infty} C(\omega) d\omega = 1.0$ to machine precision. The curve from the patching method is higher in the intermediate regime and matches the FE result in the low frequency regime. It violates the sum rule since the spectral function is obtained by approximately patching up the spectral function of each energy shell.\cite{Weichselbaum1} 

The Lehmann representation of FDM-produced $C(\omega)$ can be written as $C(\omega) = \sum_{k} w_{k}\delta(\omega - \epsilon_k)$. We separate the positive and negative components as
\begin{eqnarray}
   &&   C(\omega) = C^{(+)}(\omega) + C^{(-)}(\omega)   \nonumber \\
   &&   C^{(+)}(\omega) = \sum_{k \,\, (w_k \geqslant 0 ) } w_k \delta(\omega - \epsilon_k)  \nonumber \\
      &&   C^{(-)}(\omega) = \sum_{k \,\, (w_k < 0 ) } w_k \delta(\omega - \epsilon_k).
\end{eqnarray}    \label{Eq.17}
Here, $w_k$ and $\epsilon_k$ are the weight and energy of the $k$-th pole in $C(\omega)$, respectively.
In Fig.2(b), we compare $C^{(+)}(\omega)$, $|C^{(-)}(\omega)|$, and $C(\omega)$. In a wide frequency range including where FDM agrees well with FE, $C^{(+)}(\omega)$ and $|C^{(-)}(\omega)|$ are larger than $C(\omega)$, showing that a cancellation of errors occurs in the FDM-produced $C(\omega)$. In contrast, FE produces $C^{(-)}(\omega)=0$ at machine precision for all parameters.

We find that it is easier for the FDM-produced $C(\omega)$ to become negative when smaller $\Lambda$ and smaller broadening parameter $B$ are used. In contrast, using larger $\Lambda$ and $B$ can recover a positive $C(\omega)$ which is in quantitative agreement with the FE result, even though $C^{(-)}(\omega)$ is still present. Fig.3 compares the result of $C(\omega)$ from FDM and FE, obtained at the same parameters as in Fig.2 except for using larger $\Lambda=4.0$ and broadening parameter $B=1.2$. Although the FDM curve still contains significant negative contribution $C^{(-)}(\omega)$, the full curve $C(\omega)$ becomes positive in all frequencies and agrees quite well with that of FE. This is achieved, however, at the expense of introducing larger logarithmic discretization error and larger broadening error.

\begin{figure}[t!]
\vspace{-5.0cm}
\begin{center}
\includegraphics[width=6.2in, height=4.2in, angle=0]{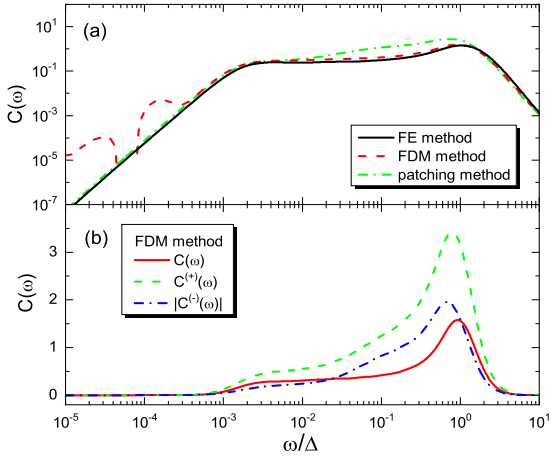}
\vspace*{-1.0cm}
\end{center}
\caption{(a) $C(\omega)$ from FE (solid line), FDM (dashed line), and patching method (dash-dotted line). (b) FDM result for $C(\omega)$, $C^{(+)}(\omega)$, and $|C^{(-)}(\omega)|$. Parameters are $s=0.3$, $\Delta=0.1$, $\alpha=0.045 > \alpha_c$, $\epsilon=0.01$, $\Lambda=2.0$, $M=200$, $N_b=12$, $B=0.5$. $T=0.0$ for the patching method and $T=10^{-8}\Delta$ for the other two.} \label{Fig2}
\end{figure}

%
\begin{figure}[t!]
\vspace{-5.0cm}
\begin{center}
\includegraphics[width=6.2in, height=4.1in, angle=0]{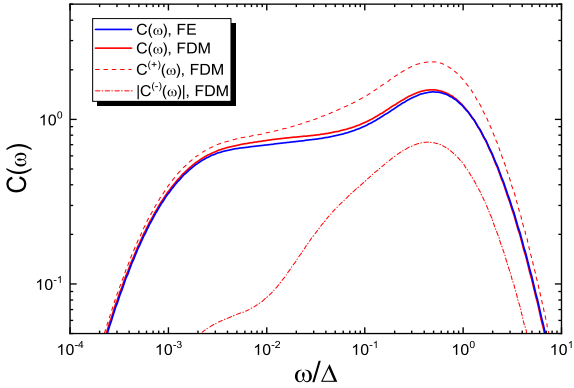}
\vspace*{-1.0cm}
\end{center}
\caption{ Compasison of $C(\omega)$ from FDM and FE for $\Lambda=4.0$ and $B=1.2$. Other parameters are same as in Fig.2.}     \label{Fig3}
\end{figure}

%
\begin{figure}[t!]
\vspace{-4.5cm}
\begin{center}
\includegraphics[width=7.0in, height=4.5in, angle=0]{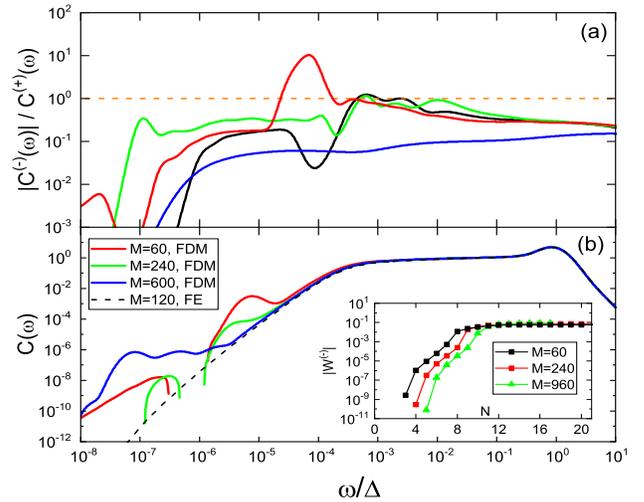}
\vspace*{-1.0cm}
\end{center}
\caption{(a) The curves of $|C^{(-)}(\omega)|/C^{(+)}(\omega)$ obtained from FDM NRG at various parameters. The horizontal dashed line marks $1.0$ above which $C(\omega)$ becomes negative. Black line: $\Lambda=2.0$, $M=80$, $N_b=6$, $B=0.5$; red line: $\Lambda=2.0$, $M=240$, $N_b=6$, $B=0.5$; green line: $\Lambda=2.0$, $M=80$, $N_b=12$, $B=0.5$; blue line: $\Lambda=4.0$, $M=240$, $N_b=6$, $B=1.2$. Other parameters are $T=10^{-8}\Delta$, $s=0.8$, $\Delta=0.1$, $\alpha=0.5 > \alpha_c$, and $\epsilon=10^{-3}$; (b) $C(\omega)$ for different $M$ values. Parameters are $T=10^{-8}\Delta$, $s=0.3$, $\Delta=0.08$, $\alpha=0.02$, $\epsilon=10^{-2}$, $\Lambda=2.0$, $N_b=6$, and $B=0.5$. Inset: integrated negative weight $|W^{(-)}|$ as functions of chain length $N$. Parameters are same as in (b) excpet $N_b=4$. }     \label{Fig4}
\end{figure}

%
\begin{figure}[t!]
\vspace{-3.8cm}
\begin{center}
\includegraphics[width=6.4in, height=4.6in, angle=0]{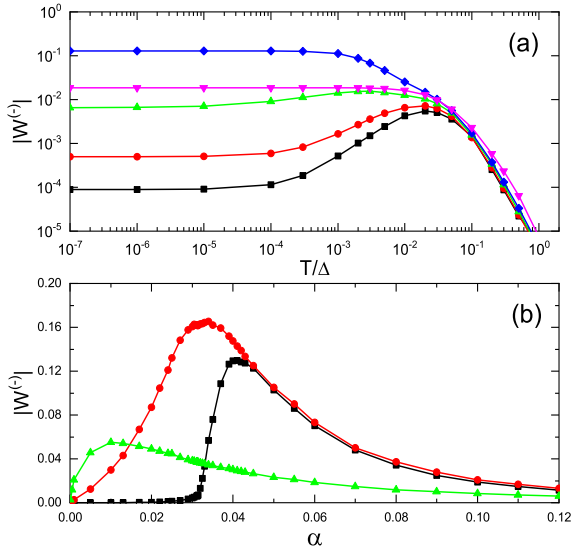}
\vspace*{-1.0cm}
\end{center}
\caption{ The absolute value of the FDM-generated negative weight, $|W^{(-)}| \equiv |\int_0^{+\infty}C^{(-)}(\omega) d\omega|$, obtained by weight summation. (a) $|W^{(-)}|$ as functions of $T/\Delta$ at $\epsilon=0.0$. From top to bottom, $\alpha=0.04$, $0.1$, $0.03128 \approx \alpha_c$, $0.02$, and $0.01$; (b) $|W^{(-)}|$ as functions of $\alpha$ at $T=10^{-8}\Delta$ and $\epsilon = 0.0$ (squares), $0.01$ (circles), and $0.1$ (up triangles). The lines are for guiding eyes. Other parameters are $S=0.3$, $\Delta =0.08$, $\Lambda=2.0$, $M=200$, and $N_b=6$.          }            
\label{Fig5}
\end{figure}

We explore whether the negative $C(\omega)$ of FDM NRG appears commonly or accidentally only at special NRG parameters. Fig.4(a) shows $|C^{(-)}(\omega)|/C^{(+)}(\omega)$ obtained from FDM method for $s=0.8$ at a low temperature $T=10^{-8}\Delta$, in the localized phase $\alpha > \alpha_c$, and with a finite bias. They are obtained using  several different combinations of NRG parameters $\Lambda$, $M$, $N_b$, and the broadening parameter $B$. The curves show that $C(\omega)<0$ (i.e., $|C^{(-)}(\omega)|/C^{(+)}(\omega)>1$) appears quite generally, especially when smaller $\Lambda$ ($\Lambda \leq 2.0$) and broadening parameter $B$ are used. This problem is not remedied by increasing $M$ and $N_b$. The only curve of $C(\omega)>0$ is obtained at $s=0.8$, using $\Lambda=4.0$ and a relatively larger broadening parameter $B=1.2$. In agreement with Fig.3, this shows that the positive $C(\omega)$ is obtained through a more effective cancellation of errors between $C^{(-)}(\omega)$ and $C^{(+)}(\omega)$ at larger $\Lambda$ and $B$. Our data for $s=0.3$ (not shown) gives the same conclusion.

FDM method should produce the exact $C(\omega)$ of $\tilde{H}_{N}(\Lambda, M=\infty, N_b)$ at $M=\infty$. We expect that the negative weight problem in $C(\omega)$ will disappear at sufficiently large $M$. Fig.4(b) shows how the FDM-produced $C(\omega)$ evolves with increasing $M$. The high frequency regime of $C(\omega)$ ($\omega/\Delta > 10^{-4}$) converges already for $M=60$. With increasing $M$, the frequency regime with converged $C(\omega)$ extends slowly towards lower frequency. The converged part agrees well with the FE curve. However, we find that the integrated negative weight $|W^{(-)}|$ does not decrease with increasing $M$. To understand this observation, using a smaller $N_b=4$, we show in the inset $|W^{(-)}|$ as functions of chain length $N$ for different $M$. For a fixed $M$, $|W^{(-)}|$ is zero for the short chain whose states are all kept. As $N$ increases further, $|W^{(-)}|$ first increases exponentially and then saturates when $N \gg \ln{M}/ln{N_b}$. For larger $M$ value, the whole curve shifts to the right but the saturated value of $|W^{(-)}|$ does not decrease. This means that in FDM, the negative weight vanishes only when $M$ covers the whole Hilbert space of the chain, i.e., when $M \gg N_b^{N}$.

In Fig.5, we explore how the FDM-produced negative weight, $|W^{(-)}| \equiv |\int_0^{+\infty}C^{(-)}(\omega) d\omega|$, changes with physical parameters $\alpha$, $\epsilon$, and $T$. Fig.5(a) and Fig.5(b) show that $|W^{(-)}|$ is largest in the parameter regime $\alpha \gtrsim \alpha_c$ and intermediate $\epsilon$. For $\alpha \ll \alpha_c$, $|W^{(-)}|$ is larger in intermediate $T$. For $\alpha \gtrsim \alpha_c$, it is larger in low $T$. We also find that the smaller $s$ is, the larger $|W^{(-)}|$ is. For $s=0.3$ shown in Fig.5(b), $|W^{(-)}|$ could be as large as $0.16$, a significant portion of the total weight of the regular part of $C(\omega)$, considering the sum rule $\int_{-\infty}^{\infty} C(\omega) d\omega =1.0$ and that $C(\omega)$ contains $c\delta(\omega)$ with $c > 0$ for $\langle \sigma_z \rangle \neq 0$. Note that $|W^{(-)}|>0$ does not necessarily imply that $C(\omega)$ becomes negative or it deviates significantly from the FE curve, because the errors in $C^{(-)}(\omega)$ and $C^{(+)}(\omega)$ may cancel each other quite accurately in $C(\omega)$ as demonstrated in Fig.3.

Now we show an example in which the negative $C(\omega)$ obtained from FDM method hinders the observation of physical phenomenon. Fig.6 shows $C(\omega)$ curves for several $\alpha > \alpha_c$ values at a finite temperature $T/\Delta =10^{-5}$. So far, the spectral function of SBM has not been studied in detail in this parameter regime. We plot both FDM and FE results. From the FE curves (solid lines), one can find a narrow frequency window around $\omega \gtrsim T$ where $C(\omega) \sim \omega^{s}$ occurs (dot-dashed straight eye-guiding lines). Below an $\alpha$-dependent low frequency $\omega_r$, $C(\omega)$ increases sharply. In the range $\omega_r \lesssim \omega \lesssim T$, a pseudo-gap forms. As $T$ decreases (not shown here), the lower boundary of this pseudo-gap range shifts towards lower frequency, forming an extended range with $C(\omega) \sim \omega^s$ behavior. In the limit $T=0$, the expected Shiba relation for the symmetry broken phase \cite{Zheng1} will be recovered. In contrast, FDM method produces negative or irregular curve (dashed lines) in the $C(\omega) \sim \omega^s$ range and the pseudo-gap range, failing to give the complete scenario.

\begin{figure}[t!]
\vspace{-2.5cm}
\begin{center}
\includegraphics[width=4.3in, height=2.9in, angle=0]{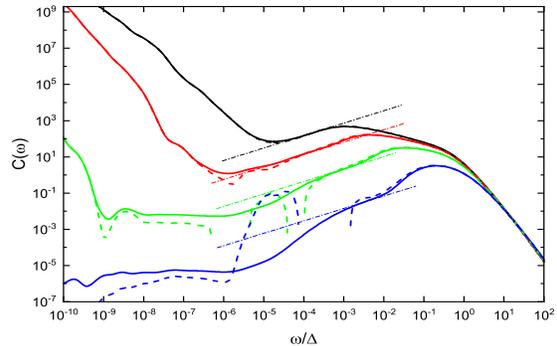}
\vspace*{-1.0cm}
\end{center}
\caption{$C(\omega)$ from FE (solid lines) and FDM (dashed lines) for $\alpha=0.117 > \alpha_c$, $0.12$, $0.13$, and $0.16$ (from top to bottom). Other parameters are $s=0.7$, $\Delta=0.01$, $\epsilon=0.0$, $T=10^{-5}\Delta$, $\Lambda=2.0$, $M=200$, $N_b=8$, and $B=0.5$. Dot-dashed straight lines are eye-guiding lines $y=cx^{0.7}$.}
\label{Fig6}
\end{figure}

%
\begin{figure}[t!]
\vspace*{-3.5cm}
\begin{center}
\includegraphics[width=5.0in, height=4.3in, angle=0]{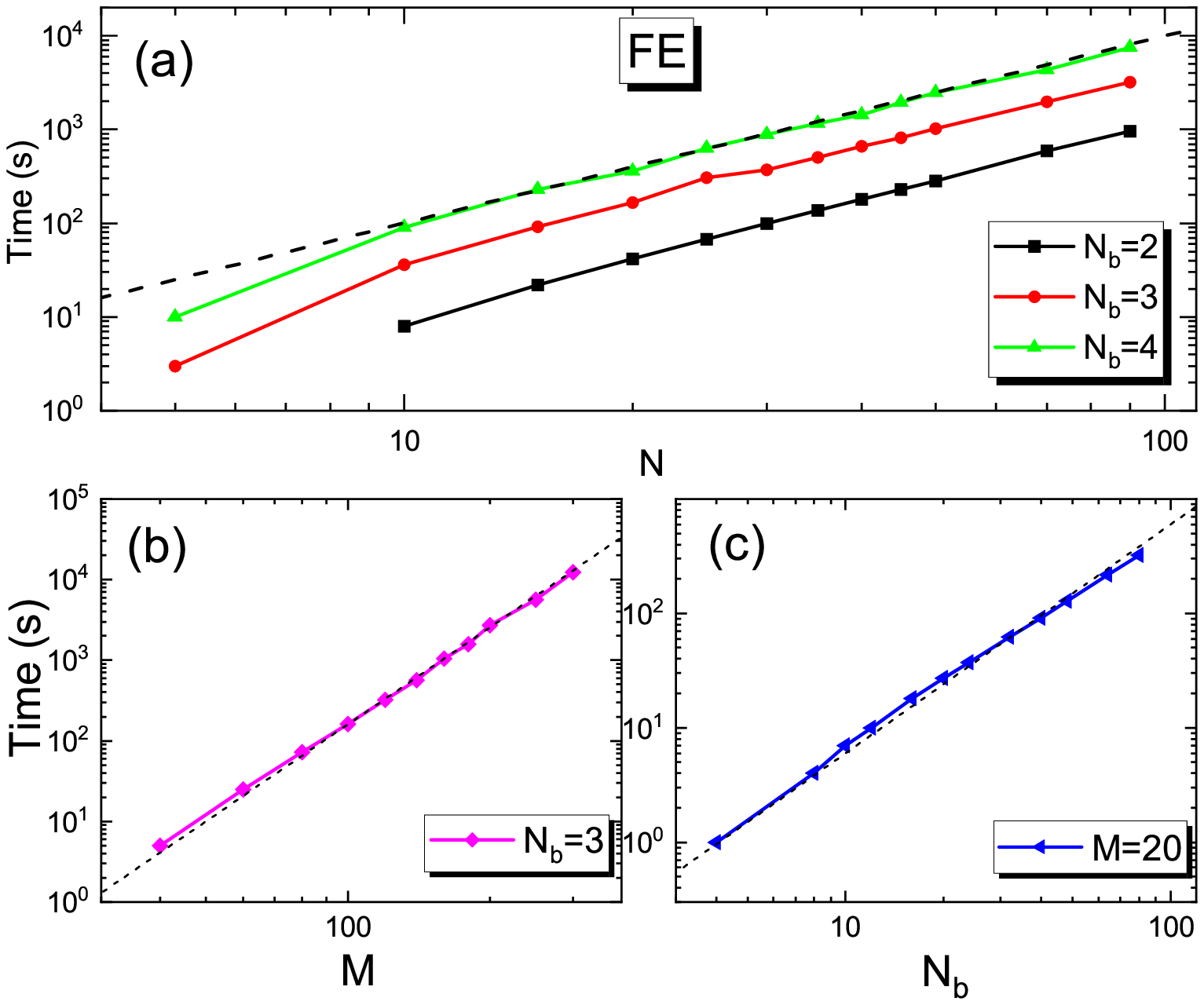}
\vspace*{-1.0cm}
\end{center}
\caption{ Scaling of computing time of FE method with respect to parameters (a) $N$; (b) $M$; and (c) $N_b$. The dashed lines are fitting lines of the form $y=c x^2$ in (a) and (c), and $y= c x^4$ in (b). Data are obtained on a single core of Intel Xeon E5-2670 with 2.6GHz frequency.
}   \label{Fig7}
\end{figure}
\begin{figure}[t!]
\vspace*{-3.8cm}
\begin{center}
\includegraphics[width=5.4in, height=4.6in, angle=0]{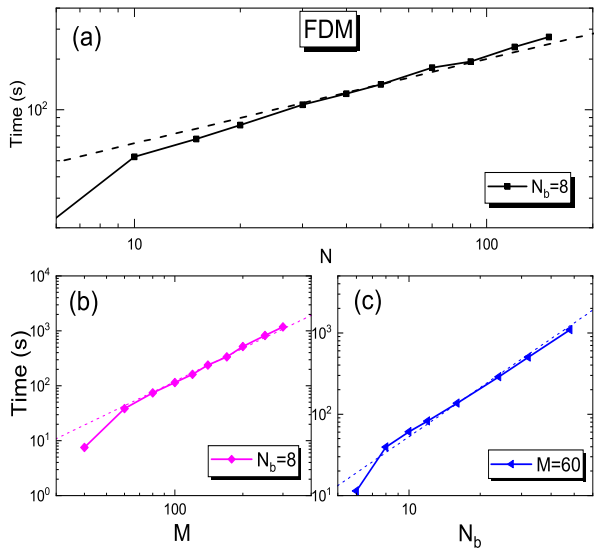}
\vspace*{-1.0cm}
\end{center}
\caption{ Scaling of computing time of the FDM method with respect to parameters (a) $N$; (b) $M$; and (c) $N_b$. The dashed lines are fitting lines of the form  $y=c x^{1/2}$ in (a), and $y=c x^2$ in (b) and (c). Data are obtained on a single core of Intel Xeon E5-2670 with 2.6GHz frequency.
}   \label{Fig8}
\end{figure}

Finally, we investigate the computing time of the FDM and FE methods. Fig.7 and Fig.8 show the scaling of the computation time of FE (Fig.7) and FDM (Fig.8) with respect to NRG parameters $N$, $M$, and $N_b$. They show that FE is more computationally demanding, with computing time proportional $M^4 N_b^{2} N^2$, while FDM method is much faster, with the scaling $M^2 N_b^2 N^{1/2}$. This is expected because the FE formalism of GF includes inter-shell excitations while the FDM one includes only intra-shell excitations. It is an open question how to modify the FE method to accelerate the computation while keep its advantages of positiveness. Both FE and FDM algorithms can be implemented with efficient parallel computing.

\section{Discussion and Summary}

The time-dependent NRG \cite{Anders1} employs the same NRG approximation as FDM does and the unitary quantum evolution is not treated accurately. Similar to the equilibrium situation, the exact result for $\tilde{H}_{N}$ will provide well-behaved time evolution of $\langle O(t) \rangle$. Therefore, the present FE for equilibrium state can be extended to time-dependent NRG for studying quantum quench problems. A comparison study will shed light on to what extent the FE method can improve the result for non-equilibrium time evolution of interested quantities.

The concept of exactly solvable effective projected Hamiltonian, such as the $\tilde{H}_{N}$ in the present work, can also be extended to other algorithms.  The energy-based truncating criterion used in ordinary NRG algorithm does not produce the optimal matrix product eigenstates. By replacing the energy-based truncating criterion of NRG with the density matrix-based criterion, or using the variational scheme of matrix product states,\cite{Weichselbaum3} NRG algorithm can be improved and a bridge between NRG and density matrix renormalization group (DMRG) has been established.\cite{Saberi1,Weichselbaum3,Pizorn1} The idea of FE could also be applied to these new NRG algorithms for better precision. For the one-dimensional quantum many-body systems with short-range entanglement, it is an interesting open question whether the exactly solvable effective projected Hamiltonian like $\tilde{H}_{N}$ can be constructed and an accurate full spectrum algorithm for the dynamical quantities can be developed.

In summary, we propose the FE algorithm for calculating the dynamical quantities of quantum impurity models in the equilibrium state. This algorithm is based on the exact solution of the projected NRG Hamiltonian and hence it circumvents the negative spectral function problem of FDM NRG. We demonstrate the effect of FE and its advantage over FDM method by a comparison study of $C(\omega)$ for SBM.

\section{Acknowledgments}
This work is supported by 973 Program of China (2012CB921704), NSFC grants (11374362, 11974420), Fundamental Research Funds for the Central Universities, and the Research Funds of Renmin University of China 15XNLQ03.

\appendix{}

\section{Derivation of FE Formalism for $G^{f/b}_{A,B}(\omega)$}

In this appendix, we derive the FE formalism for the retarded GF $G^{f/b}_{A,B}(\omega)$ defined in the main text. We use the notations in Ref. \onlinecite{Weichselbaum1}. For completeness, we also summarize the basic formulas about the complete basis developed there.  

We denote the recursive relation of NRG Hamiltonian as $H_n = H_{n-1} + \Delta H_{n-1}$. $H_N$ is the Hamiltonian of the full chain with length $N$. $H_{n_0}$ is Hamiltonian of the longest chain whose eigenstates are all kept. In the NRG iteration, after diagonalizing $H_n$, we obtain its eigenstates and eigenenergies, which are denoted as $|s \rangle_n^{X}$ and $E^{X}_{ns}$ (Ref.\onlinecite{Note1}), respectively. Here $X=D$ for the discarded states and $X=K$ for the kept states. According to NRG algorithm, there is the following recursive relation between $|s \rangle_n^{X}$ and $|s \rangle_{n-1}^{X}$,
\begin{equation}
    |s^{\prime} \rangle_n^{X} = \sum_{\sigma_n, s} |\sigma_n \rangle \otimes |s\rangle_{n-1}^{K} \left[ U^{(\sigma_n)}_{KX} \right]_{ss^{\prime}}.     \label{Eq.A1}
\end{equation}
Here, $\{ |\sigma_n \rangle \}$ ($n=1,2,...,d$) are the local states of the $n$-th chain site and $d$ is the dimension of the local Hilbert space of bath site $n$. $\left[ U^{(\sigma_n)}_{KX} \right]$ is the $K$-$X$ block of the unitary transformation matrix used to diagonalize $H_n$, which has been written in the matrix product state representation.
The orthonormal relation for a single shell, ${}^{X}_{m}\langle s | s^{\prime} \rangle^{X^{\prime}}_{m} = \delta_{XX^{\prime}} \delta_{ss^{\prime}}$, gives
\begin{equation}
   \sum_{\sigma_m} \left[ U^{(\sigma_m)}_{KX} \right]^{\dagger} \left[ U^{(\sigma_m)}_{KX^{\prime}} \right] = \mathbf{1} \delta_{XX^{\prime}}.     \label{Eq.A2}
\end{equation}

According to Ref.\onlinecite{Anders1}, a complete orthonormal basis for the full NRG chain Hamiltonian $H_N$ can be constructed by the discarded states $|s \rangle^{D}_{n}$ and the environment states $|e_n \rangle$,
\begin{eqnarray}
   |s e \rangle^{D}_{n} &=& |e_n \rangle \otimes | s \rangle^{D}_n    \nonumber \\
              &=& |\sigma_N \sigma_{N-1} ... \sigma_{n+1} \rangle \otimes |s\rangle^{D}_n.        \label{Eq.A3}
\end{eqnarray}
For the last chain site $n=N$, all the eigenstates of $H_N$ are regarded as discarded. Similarly, one can construct the
kept states $\{|s e \rangle^{K}_{n} \}$ but they do not form complete orthonormal basis.
These states have the following properties.\cite{Weichselbaum1}

(i) Orthonormal relation.
For the same shell $n=m$,
\begin{equation}
  {}^{X^{\prime}}_{n} \langle s^{\prime} e^{\prime} | s e \rangle^{X}_{n} = \delta_{XX^{\prime}} \delta_{ss^{\prime}} \delta_{ee^{\prime}} \,\,\,\,\,\,\,\,\, (X, X^{\prime} \in \{K, D \})  .          \label{Eq.A4}
\end{equation}
For different shell $n < m$,
\begin{equation}
   {}^{D}_{n} \langle s e | s^{\prime} e^{\prime} \rangle^{X}_{m} = 0  \,\,\,\,\,\,\,\,\, (X=K, D).  \label{Eq.A5}
\end{equation}

(ii) Inner product.
For $n < m$, 
\begin{eqnarray}
&&   {}^{K}_{n} \langle s e | s^{\prime} e^{\prime} \rangle^{X}_{m} = \delta_{e_{>m}, e^{\prime}_{>m}} 
   \left[ U^{(\sigma_{n+1}^{e})}_{KK}  U^{(\sigma_{n+2}^{e})}_{KK}  ...  U^{(\sigma_{m}^{e}) }_{KX}  \right]_{ss^{\prime}}. \nonumber \\
&&                    \label{Eq.A6}
\end{eqnarray}

Here, $\delta_{e_{>m}, e^{\prime}_{>m}}$ equals to unity if $\sigma_N^{e} ... \sigma_{m+1}^{e}$ of environment $e$ equals to $\sigma_{N}^{e^{\prime}} ... \sigma_{m+1}^{e^{\prime}}$ of environment $e^{\prime}$. It equals to zero otherwise.

(iii) Complete relation.
\begin{eqnarray}
 && \sum_{n=n_0+1}^{N} \sum_{s,e} |s e\rangle^{D}_n  {}^{D}_{n} \langle s e| = \mathbf{1}^{d_0 d^{N-n_0}};  \\          \label{Eq.A7}
 && \sum_{s,e} |s e\rangle^{K}_{n_0}  {}^{K}_{n_0} \langle s e| = \mathbf{1}^{d_0 d^{N-n_0}};   \\   \label{Eq.A8}
 && \sum_{n=n_0+1}^{m} \sum_{s,e} |s e\rangle^{D}_n  {}^{D}_{n} \langle s e| + \sum_{s,e} |s e\rangle^{K}_{m}  {}^{K}_{m} \langle s e| = \mathbf{1}^{d_0 d^{N-n_0}}.  \nonumber \\             \label{Eq.A9}
 &&    
\end{eqnarray}
Here $d_0$ is the number of eigenstates of $H_{n_0}$.

In this work, we suggest the following exact relation,
(iv) Eigenstates of $\tilde{H}_{N}$.
\begin{equation}
   \tilde{H}_{N} |s e \rangle^{D}_n = E^{D}_{ns}|s e \rangle^{D}_n.        \label{Eq.A10}
\end{equation}
The NRG Hamiltonian $\tilde{H}_{N}$ here is defined in Eq.(3) of the main text. This equation, together with $\tilde{H}_{N} |s e \rangle^{K}_n \approx E^{K}_{ns}|s e \rangle^{K}_n$, was called NRG approximation in Ref.\onlinecite{Weichselbaum1}. In fact, Eq.(A10) is an exact equation while the corresponding equation for the kept states is an approximation. In the derivation of FDM formalism, both equations were used.\cite{Weichselbaum1} In this work, we only use the exact equation Eq.(A10) for FE.

We start from Lehmann representation of the Fourier transformation of $\langle A(t)B \rangle$
\begin{eqnarray}
&&   \frac{1}{i} \int_{0}^{\infty} Tr \left[ \rho A(t) B \right] e^{i(\omega + i\eta)t}dt \nonumber \\
   && = \frac{1}{Z} \sum_{m,n}e^{-\beta E_n} \frac{\langle n |A|m \rangle \langle m|B|n \rangle}{\omega + i \eta + E_n - E_m}.          \label{Eq.A11}
\end{eqnarray}
Here $Z=Tr(e^{-\beta H})$ is the partition function and $\rho = e^{-\beta H}/Z$ is the density operator.
Replacing $H$ with $\tilde{H}_{N}$ and $\{ |n\rangle \}$ with $\{ |s e \rangle^{D}_n \}$ in the above equation, 
and using Eq.(A10), we obtain
\begin{eqnarray}
&&   \frac{1}{i} \int_{0}^{\infty} Tr \left[ \rho A(t) B \right] e^{i(\omega + i\eta)t}dt  = \nonumber \\
   & & \frac{1}{Z} \sum_{m,n=n_0+1}^{N} \sum_{ss^{\prime}} \sum_{ee^{\prime}} e^{-\beta E_{ns}^{D}} \frac{ {}^{D}_{n}\langle se |A| s^{\prime}e^{\prime} \rangle^{D}_{m} {}^{D}_{m}\langle s^{\prime}e^{\prime} |B| se \rangle^{D}_{n}}{\omega + i \eta + E^{D}_{ns} - E^{D}_{m s^{\prime}}}.  \nonumber \\
   &&         \label{Eq.A12}
\end{eqnarray}

For the matrix element ${}^{D}_{n}\langle se |A| s^{\prime}e^{\prime} \rangle^{D}_{m}$ with $m>n$, inserting Eq.(A9) into the right-hand side of $A$ and using Eq.(A5), we obtain
\begin{equation}
{}^{D}_{n}\langle se |A| s^{\prime}e^{\prime} \rangle^{D}_{m}  = \sum_{\tilde{s}\tilde{e}} {}^{D}_{n}\langle se |A| \tilde{s} \tilde{e} \rangle^{K}_{n} {}^{K}_{n} \langle \tilde{s} \tilde{e}|s^{\prime}e^{\prime} \rangle^{D}_{m} \,\,\,\,\,\,\,(m>n).        \label{Eq.A13}
\end{equation}
Using Eq.(A6) and 
${}^{D}_{n}\langle se |A| \tilde{s} \tilde{e} \rangle^{K}_{n} = {}^{D}_{n}\langle s |A| \tilde{s}  \rangle^{K}_{n} \delta_{e,\tilde{e}}$, we further obtain for $m > n$
\begin{eqnarray}
&&  {}^{D}_{n}\langle se |A| s^{\prime}e^{\prime} \rangle^{D}_{m} \nonumber \\
 &=& \sum_{\tilde{s}} \left[ U^{(\sigma_{n+1}^{e})}_{KK}  U^{(\sigma_{n+2}^{e})}_{KK} ... U^{(\sigma_{m}^{e})}_{KD} \right]_{\tilde{s}s^{\prime}} {}^{D}_{n}\langle s |A| \tilde{s} \rangle^{K}_{n} \delta_{e_{>m},e^{\prime}_{>m}}.  \nonumber \\
 &&          \label{Eq.A14}
\end{eqnarray}
Here, we have assumed that $A$ is a local operator defined in the impurity Hilbert space.
${}^{D}_{m}\langle s^{\prime}e^{\prime} |B| s e \rangle^{D}_{n}$ ($m>n$) can be obtained similarly. We then obtain the nominator of Eq.(A12) for $m>n$ as
\begin{eqnarray}
&& \sum_{e,e^{\prime}} {}^{D}_{n}\langle se |A| s^{\prime}e^{\prime} \rangle^{D}_{m} {}^{D}_{m}\langle s^{\prime}e^{\prime} |B| s e \rangle^{D}_{n}  \nonumber \\
&=& d^{N-m} \sum_{\sigma_{n+1} ... \sigma_m} \sum_{s^{\prime \prime}, \tilde{s}} {}^{D}_{n}\langle s |A| s^{\prime\prime} \rangle^{K}_{n}  {}^{K}_{n}\langle \tilde{s} |B| s \rangle^{D}_{n}  \nonumber \\
&& \times \left[ T_{KD}^{(nm)} \right]_{s^{\prime\prime}s^{\prime}} \left[T_{KD}^{(nm)} \right]^{\dagger}_{s^{\prime} \tilde{s}}   \,\,\,\,\,\,\,\,\,\,\, (m>n).          \label{Eq.A15}
\end{eqnarray}
In the above equation, $T_{KD}^{(nm)} = U_{KK}^{(\sigma_{n+1})} U_{KK}^{(\sigma_{n+2})} ... U_{KD}^{(\sigma_{m})}$.
The sum over environmental indices $\sigma_i$ ($i=n+1, n+2, ..., m$) contains exponentially large number of terms. We carry out this summation efficiently using the recursive formula Eqs.(12)-(14) of the main text.

The expression for $m< n$ can be obtained from Eq.(A15) by using the exchange $m \leftrightarrow n$, $s \leftrightarrow s^{\prime}$, $e \leftrightarrow e^{\prime}$, $A \leftrightarrow A^{\dagger}$, $B \leftrightarrow B^{\dagger}$ and taking complex conjugate. We split the summation $\sum_{m,n}$ in Eq.(A12) into those for $m=n$, $m>n$, and $m<n$. Inserting the respective expressions and after some simplification, we obtain
\begin{eqnarray}
 &&   \frac{1}{i} \int_{0}^{\infty} Tr \left[ \rho A(t) B \right] e^{i(\omega + i\eta)t}dt \nonumber \\
 &=& \frac{1}{Z}\sum_{n=n_0+1}^{N} \sum_{ss^{\prime}} d^{N-n} e^{-\beta E^{D}_{ns}} \frac{ \left[ B_{DD}^{(n)} \right]_{s^{\prime}s} \left[ A_{DD}^{(n)} \right]_{ss^{\prime}} }{\omega + i\eta + E^{D}_{ns} - E^{D}_{ns^{\prime}}}  \nonumber \\
 &+& \frac{1}{Z}\sum_{n=n_0+1}^{N}\sum_{m=n+1}^{N} \sum_{ss^{\prime}}  \frac{ d^{N-m} e^{-\beta E^{D}_{ns}} \left[AB \right]_{ss^{\prime}}^{(nm)} }{\omega + i\eta + E^{D}_{ns} - E^{D}_{ms^{\prime}}}  \nonumber \\
 &+& \frac{1}{Z}\sum_{n=n_0+1}^{N}\sum_{m=n+1}^{N} \sum_{ss^{\prime}}  \frac{ d^{N-m} e^{-\beta E^{D}_{ms^{\prime}}}\left[BA\right]_{ss^{\prime}}^{(nm)} }{\omega + i\eta + E^{D}_{ms^{\prime}} - E^{D}_{ns}} . \nonumber \\
 &&         \label{Eq.A16}
\end{eqnarray}
Here, $\left[ A_{DD}^{(n)} \right]_{ss^{\prime}}$ and $\left[ B_{DD}^{(n)} \right]_{s^{\prime}s}$ are defined below Eq.(9) of the main text. $\left[AB \right]_{ss^{\prime}}^{(nm)}$ and $\left[BA\right]_{ss^{\prime}}^{(nm)}$ are given by Eqs.(12)-(14) of the main text.
Eq.(A16) gives the particle part of the retarded GF. The hole part $(1/i)\int_{0}^{\infty} Tr \left[ \rho B A(t) \right] e^{i(\omega + i\eta)t}dt$ can be obtained similarly. From them, one obtains the full expression for $G^{f/b}_{AB}(\omega)$, i.e., Eqs.(10)-(14) of the main text. One can estimate that the FE computation time for GF scales as $N^2 M^4$. Parallel computation can be easily implemented for this formalism.


\end{document}